\documentclass[12pt]{iopart}
\usepackage{iopams}
\usepackage{epsfig}

\begin{document}

\title[Gersborg-Hansen {\it et al.}: Finite-element simulation of cavity modes in ...]{Finite-element simulation of cavity modes in a micro-fluidic dye ring laser}

\author{M. Gersborg-Hansen, S. Balslev, and N.~A. Mortensen\footnote[3]{Corresponding author:
nam@mic.dtu.dk}}

\address{NanoDTU, MIC -- Department of Micro and Nanotechnology, Technical University of Denmark, DK-2800 Kongens Lyngby, Denmark}

\begin{abstract}
We consider a recently reported micro-fluidic dye ring laser and
study the full wave nature of TE modes in the cavity by means of
finite-element simulations. The resonance wave-patterns of the
cavity modes support a ray-tracing view and we are also able to
explain the spectrum in terms of standing waves with a mode
spacing $\delta k = 2\pi/L_{\rm eff}$ where $L_{\rm eff}$ is the
effective optical path length in the cavity.
\end{abstract}

\submitto{\JOA}

\maketitle

\section{Introduction}

Compact, efficient, and on-chip light-sources are of considerable
interest for use in lab-on-a-chip applications
\cite{Verpoorte:2003} and recently there has been an increasing
effort in realizing micro-fluidic dye lasers based on glass or
polymer~\cite{Helbo:2003,Cheng:2004,Balslev:2005,GersborgHansen:2005,Galas:2005,Vezenov:2005}.

Typically, the cavity designs rely on classical ray-tracing
arguments rather than full wave simulations. In this paper we
consider a geometry resembling that of
Refs.~\cite{Cheng:2004,GersborgHansen:2005,Galas:2005} and offer a
full wave study of the TE modes in the cavity. The resonance
wave-patterns of the cavity modes support the ray-tracing view and
we are also able to explain the mode-spacing of the spectrum in
terms of standing waves.

The paper is organized as follows: In Sec.~2 we present the
geometry, in Sec.~3 we address the mode spacing by quasi
one-dimensional considerations, in Sec.~4 we numerically solve the
wave equation, and in Sec.~5 we discuss aspects of optical gain.
Finally, in Sec. 6 conclusions are given.

\begin{figure}[b!]
\begin{center}
\epsfig{file=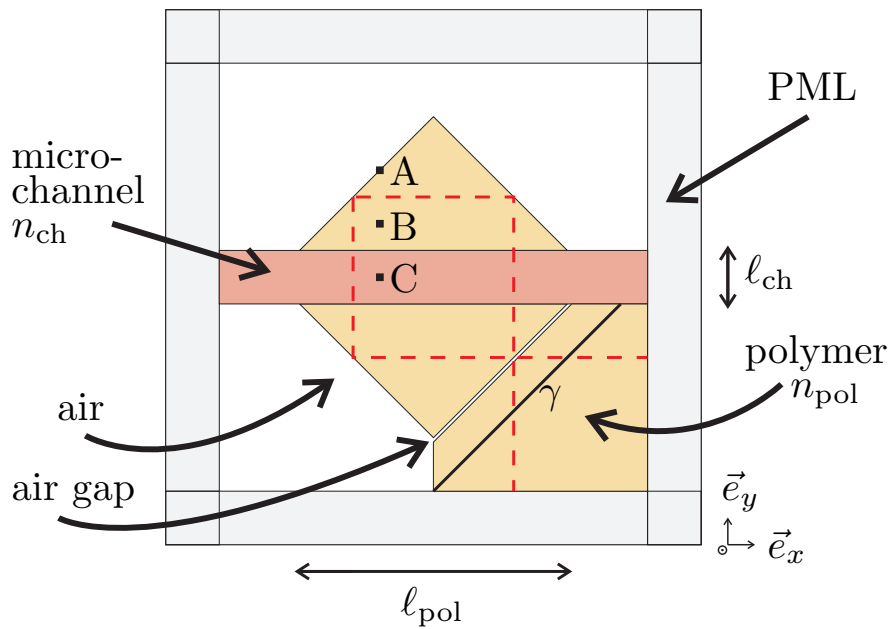,width=0.75\columnwidth,clip}
\end{center}
\caption{Geometry consisting of a polymer-defined micro-cavity
with an embedded micro-fluidic channel containing a dissolved
laser dye. Out coupling of power from the cavity occurs through an
evanescent-field coupling through an air gap to an adjacent
polymer region where the output power is evaluated by an integral
along the solid line $\gamma$. The dashed line indicates a typical
optical path in the cavity. Simulations are carried out for
point-source excitations at positions A, B, and C, respectively. }
\label{fig1}
\end{figure}

\section{Geometry}

We consider the two-dimensional laser resonator illustrated in
Fig.~\ref{fig1} which corresponds to the planar cavities studied
experimentally in
Refs.~\cite{Cheng:2004,GersborgHansen:2005,Galas:2005}. The cavity
resembles a classical Fabry–-Perot resonator and consists of two
dielectric isosceles triangles with baseline $\ell_{\rm pol}$ and
refractive index $n_{\rm pol}$ separated by a microfluidic channel
of width $\ell_{\rm ch}$ containing a fluid with refractive index
$n_{\rm ch}$. Light is confined to the cavity by total-internal
reflections at the polymer-air interfaces at an angle of incidence
of $\pi/4$. Out coupling of power occurs through an
evanescent-field coupling to an adjacent polymer region. In the
experiments in Refs.~\cite{GersborgHansen:2005,Galas:2005} the
microfluidic channel is filled by a dye doped liquid acting as
gain medium. In Ref.~\cite{GersborgHansen:2005} $\ell_{\rm
pol}\sim 700\,{\rm \mu m}$, the cavity is pumped at the wavelength
$\lambda=532\,{\rm nm}$ by a pulsed frequency doubled Nd:YAG
laser, and lasing occurs in the visible around $\lambda\sim
570\,{\rm nm}$. For details on the pump power and lasing threshold
we refer to Ref.~\cite{GersborgHansen:2005}.

Throughout the rest of the paper we consider a typical structure
with $\ell_{\rm ch}/\ell_{\rm pol}=0.2$ and for the
evanescent-field coupling we have $w/\ell_{\rm pol}\simeq 0.028$
for the width $w$ of the air gap. For the refractive indices we
use $n_{\rm pol}=1.6$ and $n_{\rm ch}=1.43$. These numbers give an
index step which is typical for a liquid and a polymer. However,
we emphasize that the particular choice of numbers do not affect
our overall findings and conclusions.

\section{Quasi one-dimensional approach to mode spacing}

We first estimate the mode spacing by considering a plane wave
travelling around in the cavity, see Fig.~\ref{fig1}. In this
ray-tracing like approach we neglect reflections at the
polymer-fluid interfaces which is justified by the very small
Fresnel reflection probability
\begin{equation}
R= \left(\frac{n_{\rm pol}-n_{\rm ch}}{n_{\rm pol}+n_{\rm
ch}}\right)^2\simeq 0.31\,\%
\end{equation}
We imagine modes somewhat similar to whispering-gallery modes
(WGMs) in resonators of circular shape. However, in this case the
modes are subject to four total-internal reflections at an
incidence angle of $\pi/4$ irrespectively of the mode-index and
all modes have the same effective optical path length. Contrary to
WGMs these modes have thus no cut-off for decreasing mode index
caused by decreasing incidence angle. The accumulated phase during
one round-trip of a plane-wave in the ring cavity is
\begin{equation}
\delta\phi=k L_{\rm eff}+\varphi
\end{equation}
where $k=2\pi/\lambda=\omega/c$ is the free-space wave number,
\begin{equation}
L_{\rm eff}=2 n_{\rm pol}\ell_{\rm pol}+2n_{\rm ch}\ell_{\rm ch}
\end{equation}
is the effective optical path length in the cavity, and
\begin{equation}
\varphi=4\times \arg\left( \frac{\cos(\frac{\pi}{4})-\sqrt{n_{\rm
pol}^{-2}-\sin^2(\frac{\pi}{4})}}{\cos(\frac{\pi}{4})+\sqrt{n_{\rm
pol}^{-2}-\sin^2(\frac{\pi}{4})}}\right)
\end{equation}
is the phase picked up during the four total-internal reflections
at the polymer-air interfaces at incidence angle of $\pi/4$. The
resonance condition is $\delta\phi=2\pi m$ with the mode-index $m$
being an integer. Obviously, the corresponding modes
\begin{equation}\label{eq:km}
k_m= \frac{2\pi m-\varphi}{L_{\rm eff}}
\end{equation}
are equally spaced with the mode-spacing
\begin{equation}
\delta k= \frac{2\pi}{L_{\rm eff}}.
\end{equation}

\begin{figure}[t!]
\begin{center}
\epsfig{file=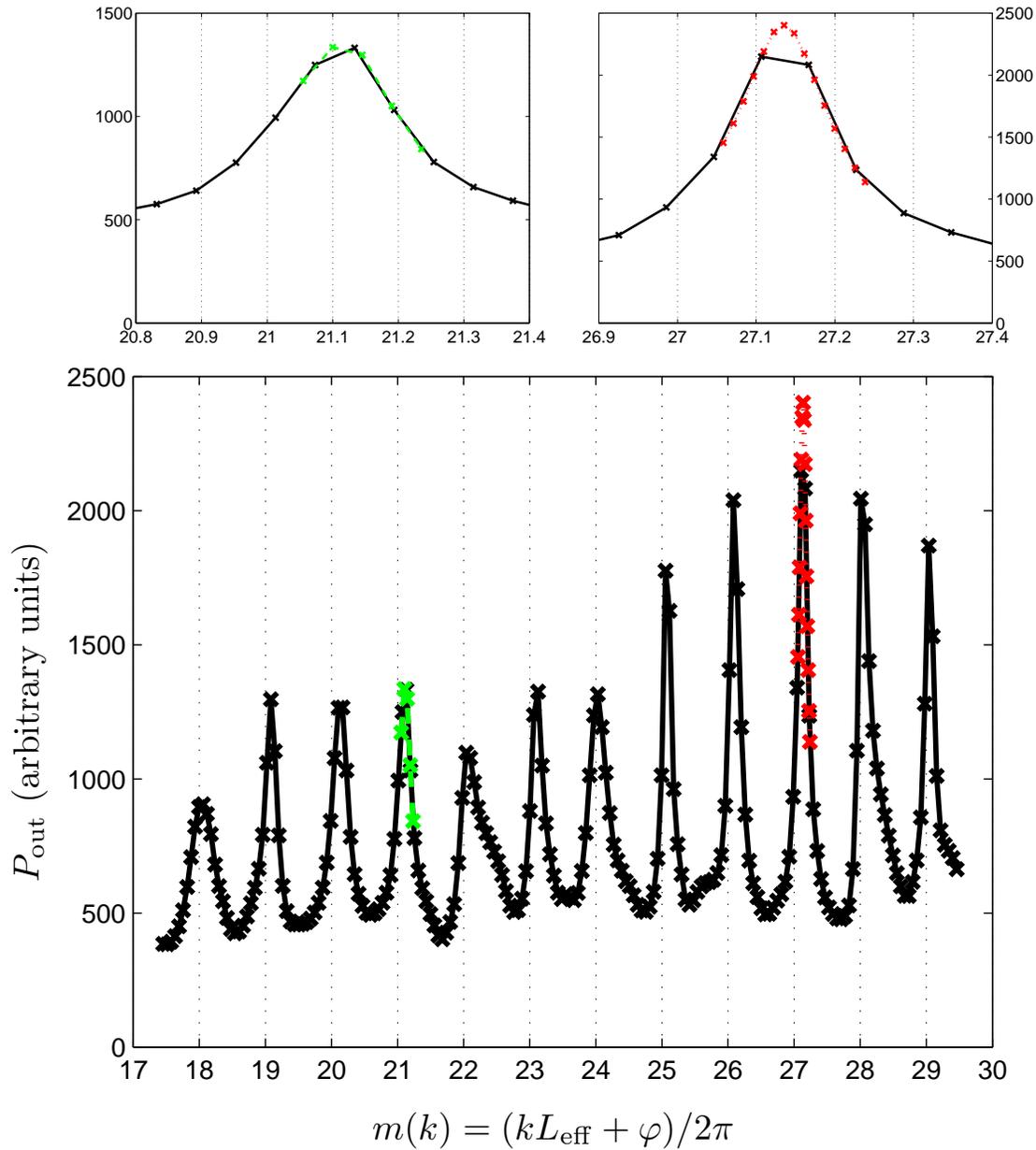,width=\columnwidth,clip}
\end{center}
\caption{Mode spectrum for a point-source excitation at position
A, see Fig.~\ref{fig1}. The top panels show close-ups with a
higher resolution of the respective peaks at $m\sim 21$ and $m\sim
27$ indicated in the lower panel.} \label{fig2}
\end{figure}

\section{Two-dimensional wave equation approach}

The full wave nature is governed by the wave
equation~\cite{Joannopoulos:1995}
\begin{equation}\label{waveequation}
\nabla\times\nabla\times
\vec{E}(\vec{r})=\epsilon(\vec{r})k^2\vec{E}(\vec{r})
\end{equation}
where $\vec{E}$ is the electrical field and
$\epsilon(\vec{r})=n^2(\vec{r})$ is the dielectric function. We
solve the wave equation in a planar geometry for TE modes, i.e.
$\vec{E}(\vec{r})=E_z(\vec{r})\vec{e}_z$ and
$\vec{r}=x\vec{e}_x+y\vec{e}_y$. For the simulations we employ a
finite-element method \cite{comsol} with "open" boundary
conditions taken into account by perfectly matching layers (PMLs)
at the edges of the simulation domain~\cite{Jin:2002}, see
Fig.~\ref{fig1}. This allows outgoing waves with negligible back
reflection.

We solve Eq.~(\ref{waveequation}) subject to a point-source
excitation and modes are monitored by calculating the output power
$P_{\rm out}(k)$ by integration along $\gamma$ in the polymer
region adjacent to the cavity, see Fig.~1, for different values of
$k$. The point-source has the appealing feature that it radiates
isotropically in a homogeneous space and thus it will in general
excite the full spectrum of cavity eigenmodes (except of course
from the statistically few having a true node at the exact
position of the point-source).

In order to compare to the predicted mode spectrum, we have
transformed the $k$ values into a mode index
\begin{equation}
m(k)=(k L_{\rm eff}+\varphi)/2\pi
\end{equation}
 and according to Eq.~(\ref{eq:km}) we expect $P_{\rm out}(m)$ to have resonances
centred at integer values of $m$. Fig.~2 illustrates this in the
case of a point-source excitation at point A, see Fig.~1.

The over-all agreement between the full wave simulation and the
quasi one-dimensional model is excellent, but from Fig.~\ref{fig2}
it is also clear that the different peaks are slightly
blue-shifted from integer values. The top panels illustrate this
for two of the peaks indicated by green and red in the lower
panel. This small shift may originate in a slightly modified phase
shift at the edge with evanescent field coupling compared to the
three other edges of the cavity. The small Fresnel reflection may
also slightly modify the spectrum compared to the results derived
from the quasi one-dimensional model.

Figure~\ref{fig2} shows results in the range from $m\sim 18$ up to
$m\sim 29$. When further increasing $m$ the pattern of peaks
persist with a small tendency that the peaks sharpen. This trend
has been investigated up $m\sim 100$ where simulations turn highly
computationally demanding (results not shown). However, since the
quasi one-dimensional interpretation does not support a cut-off
for increasing $m$ we believe that a spectrum of equally spaced
modes persist for increasing $m$.

For decreasing $m$ WGMs will typically experience a cut-off
because the angles of incidence at some point do not support
total-internal reflection. However, as discussed for the quasi
one-dimensional model the particular class of modes in the present
cavity do not share this property. In fact, in the simulations we
have observed the modes down to $m\sim 10$ below which pronounced
deviations from the quasi one-dimensional predictions start to
emerge. Deviations most likely appear because the polymer-air
interface has spatial variations on a length scale comparable to
the wavelength of the light. In other words, the ray-tracing
picture fails and concepts like total-internal reflection derived
from Snell's law do not accurately capture the true wave physics.

\begin{figure}[t!]
\begin{center}
\epsfig{figure=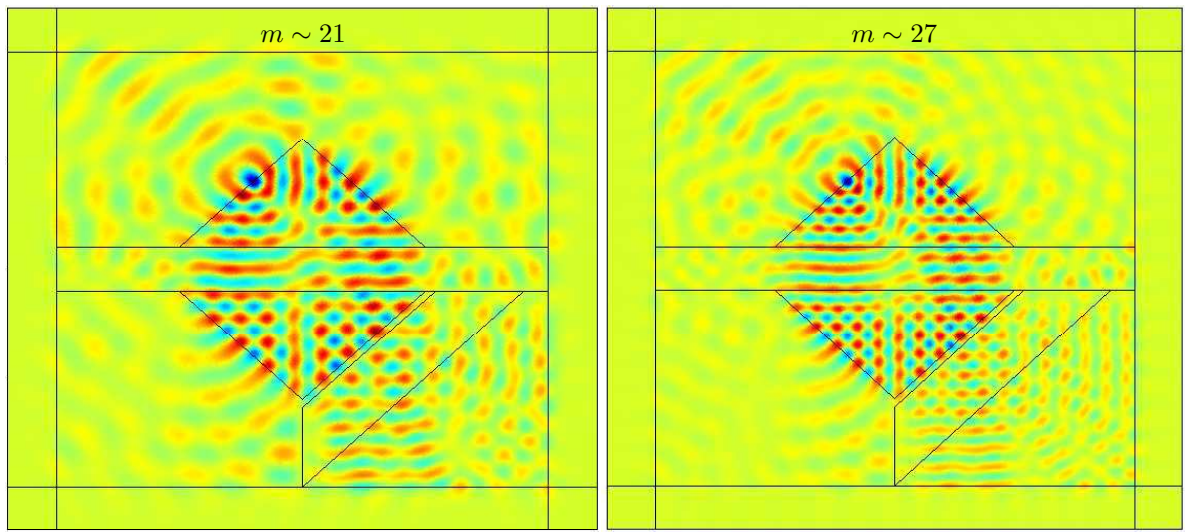, width=\textwidth,clip}
\end{center}
\caption{Electrical fields at $m(k)=21.1000$ and $m(k)=27.1355$
for a point-source excitation at position A, see Fig.~\ref{fig1}.
} \label{fig3}
\end{figure}

In order to verify that the peaks in Fig.~\ref{fig2} really do
correspond to cavity modes we have studied the corresponding
electrical fields at resonance, see Fig.~\ref{fig3}. These fields
resemble pure eigenfunctions of the resonator while off-resonance
fields correspond to linear combinations of a larger number of
eigenfunctions. Starting from e.g. the source point, the number of
oscillations along one round trip equals $m$ in full agreement
with the quasi one-dimensional arguments.

When the cavity is excited at different positions the overall
output spectrum is the same such that peaks remain unshifted while
changes are observed in the intensity distribution only. The
reason is that different positions of the source will excite
different linear combinations of eigenmodes (being correlated with
the intensity level) while the eigenspectrum itself (being
correlated with the resonance positions) remains unchanged. In
Fig.~4 we illustrate this for different positions of the point
source. The spectrum also reveals structure, though very broad
with low intensity, in between integer values of $m(k)$. This
structure also corresponds to quasi eigenmodes which however are
much more poorly confined to the cavity compared to the
well-confined modes with integer values of $m(k)$.

\begin{figure}[t!]
\begin{center}
\epsfig{file=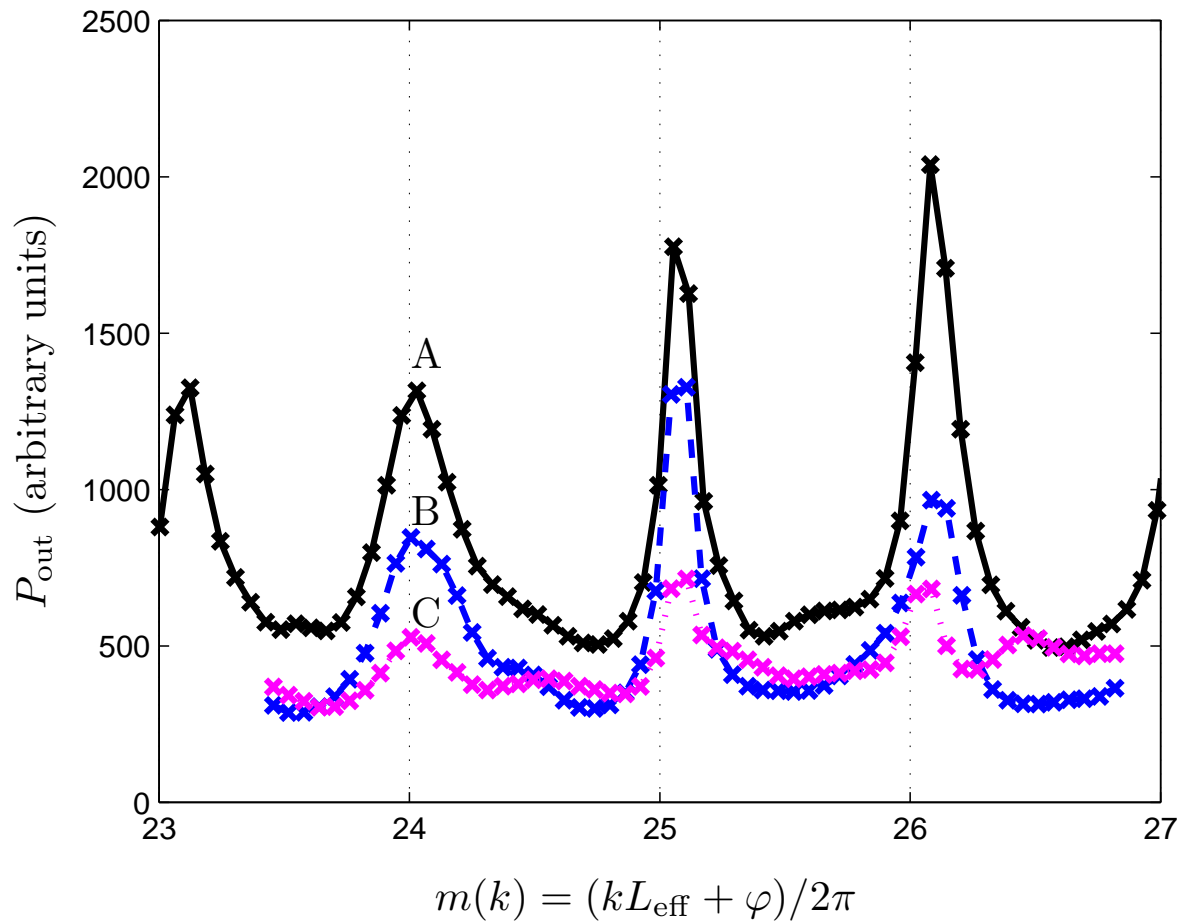,width=\columnwidth,clip}
\end{center}
\caption{Mode spectra for point-source excitation at positions A,
B, and C, see Fig.~\ref{fig1}. } \label{fig4}
\end{figure}

\section{Optical gain medium}

Lasing of course relies on the presence of an optical gain medium.
In Refs.~\cite{Cheng:2004,GersborgHansen:2005} dissolved laser dye
in the microfluidic channel provides the gain. While the dynamics
of lasing is difficult to address we may with little effort
investigate the influence of gain on the mode spectrum. The
present numerical model allows for such investigations, but for
low concentrations general trends may be more easily analyzed with
the aid of perturbation theory~\cite{Johnson:2002}. Doping by e.g.
Rhodamine 6G (Rh6G), as in the experiment
\cite{GersborgHansen:2005}, will change the refractive index in
the channel accordingly, i.e. $n_{\rm ch}\rightarrow n_{\rm ch}+
n_{\rm Rh6G}$ where for the latter $n_{\rm Rh6G}=n_{\rm Rh6G}'+i
n_{\rm Rh6G}''$. In the case of $n_{\rm Rh6G}''\ll n_{\rm
Rh6G}'\ll n_{\rm ch}$ we get
\begin{equation}
\Delta k = -\frac{k}{2} \frac{\big< \vec{E}\big|\Delta
\epsilon\big|\vec{E}\big>}{\big<
\vec{E}\big|\epsilon\big|\vec{E}\big>},\quad
\Delta\epsilon=\big(n_{\rm ch}+ n_{\rm Rh6G}\big)^2-n_{\rm
ch}^2\simeq (n_{\rm Rh6G}')^2
\end{equation}
from which we expect a red-shift of the modes of the order $\Delta
k \propto (n_{\rm Rh6G}')^2k$ along with a narrowing of the modes.
A blue-shift may be observed in the case where $n_{\rm Rh6G}''
> \sqrt{n_{\rm Rh6G}'(n_{\rm Rh6G}'+n_{\rm ch})}$.

\section{Discussion and conclusion}

In this work we have used finite-element simulations to study the
cavity mode spectrum of a micro-fluidic dye ring laser with a
planar geometry resembling the one studied experimentally in
Refs.~\cite{Cheng:2004,GersborgHansen:2005}. We have performed a
full wave study of the TE modes in the cavity and found very good
agreement with a quasi one-dimensional plane wave description with
resonances corresponding to standing waves.

In principle our simulations allow for an estimate of the quality
factor of the modes, but realistic simulations for the
experimental device require more details to be taken into account.
For instance one would need to include the three-dimensional
nature of the device to describe the radiation field accurately
and the details of the evanescent field coupling would also
influence the quality factor. Such issues add to the difficulty in
addressing the dynamics of lasing so in this work we have only
addressed the passive device. However, we have estimated a
doping-induced shift of the spectrum by perturbative means.

In the simulations we have considered mode-indices $m(k)$ up to
around 100 while in the experiments the corresponding typical mode
index is estimated to be around two orders of magnitude larger.
Nevertheless, we are confident that the standing-wave
interpretation may be safely extrapolated to the experimental
regime~\cite{Cheng:2004,GersborgHansen:2005,Galas:2005} due to the
scale invariance of the wave equation~\cite{Joannopoulos:1995} and
the fact that this class of modes has no cut-off with respect to
increasing mode index.

\section*{Acknowledgment}
We thank A. Kristensen for stimulating discussions. The work was
supported by the Danish Technical Research Council (STVF, grant
no. 26-02-0064).
\newpage


\end{document}